\documentclass[11pt]{article}

\usepackage{paithan}

\title{Games on the Sperner Triangle}
\author{Kyle W. Burke \\ Boston University \and Shang-Hua Teng\\
Boston University}

\begin{document}

\maketitle

\begin{abstract}
We create a new two-player game on the Sperner Triangle based on Sperner's lemma.  Our game has simple rules and several desirable properties.  First, the game is always certain to have a winner.  Second, like many other interesting games such as Hex and Geography, we prove that deciding whether one can win our game is a $PSPACE$-complete problem.  Third, there is an elegant balance in the game such that neither the first nor the second player always has a decisive advantage.  We provide a web-based version of the game, playable at: {\it http://cs-people.bu.edu/paithan/spernerGame/}.  In addition we propose other games, also based on fixed-point theorems.
\end{abstract}

\section{Introduction}

The relationship between computational complexity and game strategies has encouraged the development of both of these fields.  Games, due to their enjoyable and competetive nature, create a breeding pool for analysis as strategies are discussed and revised.  The ability to express strategies using computational complexity allows us to categorize them based on concrete classes of difficulty.  Inversely, the ability to express complexity classes in terms of finding game strategies motivates the study of these classes.  
%Since the complexity landscape between $P$ and $PSPACE$ remains wide open, the field still welcomes any such motivation.

Many two-player games can employ simple rules yet still resist simple methods to efficiently produce winning strategies.  Games such as Geography, Hex and Go all require $PSPACE$-complete capability to always deduce these winning strategies \cite{Papadimitriou:1994,Reisch:1981,LichtensteinSipser:1980}.  For these games, the simplicity of the rules often masks the mathematical intricacies of the underlying structure.  In Hex, for instance, the existence of a winner has been shown to be equivalent \cite{Gale:1979} to Brouwer's fixed-point theorem \cite{Brouwer:1910}.

Motivated by the equivalence of the fixed-point theorem to another result, namely Sperner's Lemma \cite{Sperner:1928}, we present a new two-player game.  As Eppstein argues, games with polynomial strategies lose their fun once players learn ``the trick''\cite{Eppstein}.  This concern might be amplified in this day and age when we encourage the most talented human game players to compete with highly optimized machines.  Therefore, if polynomial strategies for a game exists, machines can efficiently implement the strategies, and the game play becomes trivial.  Thus, we continue by proving that our game is $PSPACE$-complete.  

Furthermore, our game avoids one of the inelegant imbalances of games such as Hex, in that it is not always to a player's advantage to make the next move.  Indeed, in our game the final move results in a loss.  Thus this game does not fall into the trap of clearly giving the first person to play any advantage.  There is no need to incorporate an unnatural means to smooth this out (Hex uses the ``Pie Rule'' to attempt to undo this advantage, but it winds up giving an advantage to the second player instead).

We dub our game Sperner, in honor of the creator of the game board as well as the lemma which motivated its invention.  The rules of the game are very simple, and a player does not require any mathematical background to be a fierce contender\footnote[1]{During the recent Thanksgiving break, the first author played against his mother, who is not a mathematician, and was beaten more than once in a short span of games.}.  Before we define the rules of the game, we discuss the Sperner Triangle and the Lemma which---just as in Hex---will ensure that one of the players has won at the end of the game.

\section{The Sperner Triangle}

\begin{figure}[h]
	\centering
	\includegraphics[scale = .25]{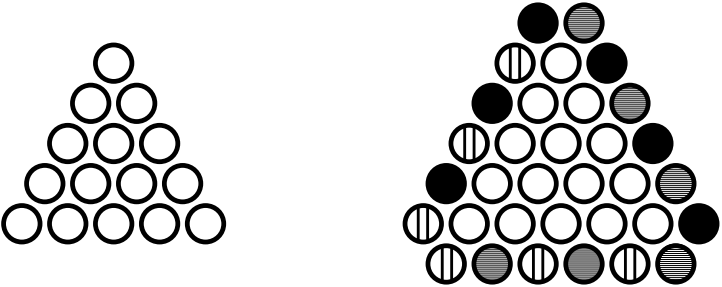}
	\caption{The original Sperner Triangle of size 5 (left) next to our Sperner Triangle gameboard, also of size 5, with added functional boundaries (right).}
	\label{fig:elegantBoundaries}
\end{figure}

Sperner's Triangle is simply a triangular array of nodes (see the left half of Figure \ref{fig:elegantBoundaries}) each colored in one of three colors with a simple boundary condition: each side of the outer triangle is assigned a different color, and nodes along that edge may not be given that color \cite{Sperner:1928}.  Along each axis of the array, each node has two natural neighbors, aside from the boundary nodes, which may not have a second neighbor along some axes.  Since the triangular array has three axes, each interior node of the triangle has exactly six neighbors.

\begin{figure}[h]
	\centering
	\includegraphics[scale = .25]{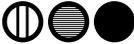}
	\caption{From left to right: A barred node, a shaded node and a filled node.}
	\label{fig:symbols}
\end{figure}

Here, in lieu of colors, we use three different symbols: bars, shading and filled-in, as demonstrated in Figure \ref{fig:symbols}.  We will often use the verbs \emph{barring}, \emph{shading} and \emph{filling} to describe the action of assigning the relative symbol to a node.

Our games are inspired by the following brilliant result of Sperner \cite{Sperner:1928}:

\begin{lemma}[Sperner's Lemma]
On any sized Sperner Triangle, if all the nodes are colored, there will exist a triangle of three neighboring nodes, each of which has a different color.  In fact, there will be an odd number of these triangles.
\end{lemma}

\section{Playing Games on the Sperner Triangle}

Given this challenge of attempting to avoid tri-colored triangles, natural games emerge in which players take turns coloring empty nodes on an initially-uncolored Sperner Triangle.  In these games, a player loses when their play creates a tri-colored triangle.  Because of this, we refer to triangles with three colors as ``bad'' triangles.  In addition, open circles at which any color played would result in a bad triangle are called ``doomed'' circles.  Although playing at one of them is legal, it results immediatley in a loss, and thus we often refer to them with the ``unplayable'' misnomer.  Also, we often speak of playable circles as not including those which are doomed.  When we want to include doomed circles, we will use the term ``open''.

\subsection{The Game Board.}

In order to elegantly enforce the border restrictions, we enhance the triangle by adding an alternating series of the other two colors to each side, as shown in the right-hand side of Figure \ref{fig:elegantBoundaries}.  Thus, if a player plays the forbidden color along one of those sides, they immediately create a bad triangle and lose the game.

Since it is often helpful to focus on one player whilst analyzing two-player games, we will often refer to one player as the hero and their opponent as the adversary.

\subsection{Sperner Rules}

\begin{itemize}

\item[1.] On their turn, each player colors a circle on the triangle one of the three colors.

\item[2.] The first player may choose any uncolored circle at which to play for the first move.

\item[3.] If any circles adjacent to the last-colored circle are uncolored, the current player must choose one of them to color.  Otherwise, they may choose to play at any uncolored circle on the board.

\item[4.] The first player to color a circle which creates a tri-colored triangle loses the game.

\end{itemize}

We call this game \emph{Sperner}.  A playable version is available at:

\begin{center}
\emph{http://cs-people.bu.edu/paithan/spernerGame/}
\end{center}

\subsection{An Alternative Game}

In this game, we follow the rules of Sperner except that players are never restricted to play adjacent to the last-colored circle.  Instead, every turn, they may choose any uncolored circle on the board.  We call this game \emph{Unrestricted Sperner}.  For more discussion concerning this game, see Section \ref{sec:unrestrictedSperner}.

\section{On The Complexity of Sperner}

Our central complexity question concerns the following decision problem:

\begin{quote}
\texttt{SPERNER}: Given a legal game state, determine whether the current player has a winning strategy.
\end{quote}

Before analyzing the complexity of this problem, we must comment on the phrase ``legal game state''.  A legal state of the game is one which can be realized from any legal sequence of plays from an initial game board.  A game state then consists of either:

\begin{itemize}

\item[1.] An initial game board.

\item[2.] A game board attainable from some sequence of moves on an initial game board, with the last move identified.

\end{itemize}

As our main complexity result, we prove:

\begin{theorem}[Main]
\texttt{SPERNER} is $PSPACE$-complete.
\end{theorem}

\subsection{Sperner is in $PSPACE$}

\begin{lemma}
\texttt{SPERNER} is in $PSPACE$.  
\end{lemma}

\begin{proof}

Since the number of plays is at most the number of nodes in the gameboard, the depth of every branch of the game tree is linear in the size of the input.  Thus, in polynomial space we can determine the result of following one path of the game tree.  In order to search for a winning result, we can systematically try each possible game branch.  Thus, we require only space enough to evaluate one branch at a time, as well as some bookkeeping to recall which branches we have already visited.  This bookkeeping will require at most only $O(m^2)$ space, where $m$ is the number of nodes on the board.  Thus, in polynomial space, we can evaluate all the possible outcomes of the game tree until we either find a winning strategy or determine that none exists.

\end{proof}

\subsection{Outline of the Reduction}

\label{sec:reductionOutline}

It remains to be shown that strategies for the Sperner Game are $PSPACE$-hard.

Classically, we show that problems are $PSPACE$-hard by reducing \texttt{TQBF} to them \cite{PapadimitriouBook:1994}.  In general, \texttt{TQBF} is the problem of determining whether a quantified boolean formula---a formula of the form $\exists x_1: \forall x_2: \exists x_3: \forall x_4: \cdots Q_nx_n: \phi(x_1, x_2, \ldots, x_n)$---is true.  In our notation here, $\phi(x_1, \ldots, x_n)$ is a conjunctive normal form formula using the literals $x_1$ through $x_n$, while $Q_n$ is a quantifier (either $\forall$ or $\exists$).

Because of the inherent alternation in quantified boolean formulae, many games with non-obvious strategies for two players have been shown to be $PSPACE$-hard \cite{PapadimitriouBook:1994}. Indeed, we see that fulfilling a \texttt{TQBF} instance is much like playing a game.  The hero will choose a variable assignment for $x_1$, then the adversary will choose for $x_2$.  The hero chooses $x_3$ in response, and so on. 

Our reduction will model this behavior.  We will create a legal Sperner game state from a \texttt{TQBF} instance such that a winning strategy in the game exists if and only if the formula is true.  Our reduction is inspired by the reduction of \texttt{TQBF} to \texttt{GEOGRAPHY} (see page 460 of \cite{PapadimitriouBook:1994} for a clear description of this reduction).  The game will proceed by letting the appropriate players make moves corresponding to the assignment of values to the variables $x_i$.  Each player will then make one further choice and one of the literals in one of the clauses will be selected.  We will ``investigate'' the literal through its interpretation in the game state and force an end of the game with it.

In our resulting game boards, most of the moves players make will be very restrictive.  In our construction, there is a resulting ``prescribed flow of play'' directing players to make choices in the order we described above.  Our reduction provides punishment strategies such that any player violating the prescribed flow of play will lose in a constant number of turns.  We describe these punishment strategies, ensuring that players must follow the prescribed flow in order to have a chance of winning the game.

Using our construction, each player's last choice is easily described: the adversary will choose a clause to investigate, and the hero will choose one of the literals in that clause.  That literal will be evaluated, according to the assignment it received.  If the literal is true, the hero should win, otherwise the adversay should win.

\begin{figure}[h]
	\centering
	\includegraphics[scale = .5]{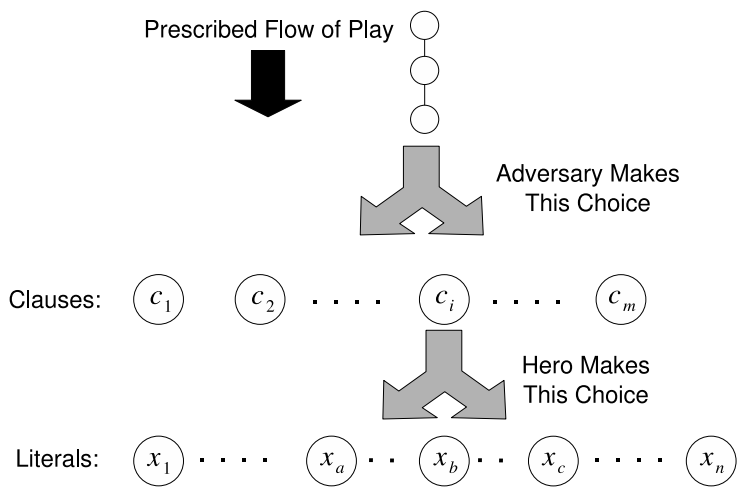}
	\caption{Construction Sketch: End of the Game.  Clause $c_i$ contains the literals $x_a, x_b, x_c$.}
	\label{fig:endOfSpernerGamePlan}
\end{figure}

We give the adversary the power to pick a clause because, in the case where the formula is true, all the clauses must be true; the adversary will have no power.  However, if at least one of the clauses is false, the formula will be false, and the adversary should be able to select one of these false clauses in order to discredit the correctness.  Conversely, inside each clause we will give the hero the ability to choose between literals.  Thus, if at least one of those literals is true, the hero will be able to choose and identify it.  Figure \ref{fig:endOfSpernerGamePlan} illustrates the layout style we desire.

\begin{figure}[h]
	\centering
	\includegraphics[scale = .5]{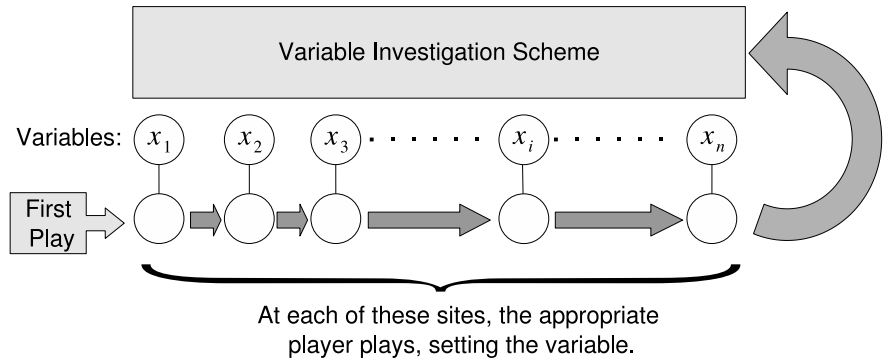}
	\caption{Construction Sketch: Setting the Variables.  The Variable Investigation Scheme is laid out in Figure \ref{fig:endOfSpernerGamePlan}}
	\label{fig:settingVariableConstructionSketch}
\end{figure}

Before the flow of play reaches this point, we must have already set all of the variables.  In order to accomplish this, the path of play must first pass by each of the variable settings, forcing the appropriate player to make a decision at each variable.  Once the settings have been accomplished, we can move to the investigation procedure, as portrayed in Figure \ref{fig:settingVariableConstructionSketch}.

Overall, this plan is very reminiscient of the reduction from GEOGRAPHY.  In addition, our topology meets some similar hurdles that must be overcome in that construction.  For instance, our plan has a non-planar design (paths must often intersect between the selection of a clause and the selection of a literal during investigation).  Thus, we will need to produce widgets which allow for these logical crossovers to occur, as we do in the following section.

Our blueprints also seem to defy the rigid structure of the gameboard we are operating on.  We require widgets which provide pathways for our prescribed flow of play.  GEOGRAPHY is played on a directed graph, so enforcing the flow of play is somewhat more simple.  We will need to be very careful that players cannot subjugate design plans by moving in unexpected directions.  Also, we need widgets to handle variable assignment, path splitting, and other obstacles to realizing our layout on a Sperner Board.  We continue by exhaustively describing these widgets.

\section{Reduction Widgets}

Our reduction requires widgets to enforce various moves and allow for appropriate decisions to be made through others.  In addition, the widgets must be able to connect, allowing us to build the overlying structure by fitting them together.  In this section, we describe each of the widgets and specify how they are used.

Many of the widgets are simple and are only pathways to guide the flow of play.  For more complex widgets, however, we need to be able to ensure that plays not following this flow correspond to trivially bad choices.  This means that any player attempting to go against the prescribed pathway will be vulnerable to an optimal opposing winning strategy which is easily computable in constant time.

\subsection{Paths}

Paths are the most simple of the widgets in our construction, although we have two different versions for different circumstances.  Players should not make non-trivial decisions along paths, thus we build them to strongly restrict playing options.

\begin{figure}[h]
	\centering
	\includegraphics[scale = .25]{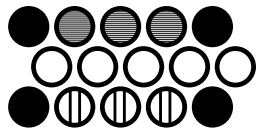}
	\caption{A Single-Symbol Path}
	\label{fig:onePathWidget}
\end{figure}

The first of our two versions is a path in which on any move a player has the option of playing exactly one symbol without immediately losing.  In Figure \ref{fig:onePathWidget}, assuming the flow of play comes in from the left, the leftmost circle can and must be filled.  Then, the next player is forced to fill the circle to the right, and so on.  This path pattern can be extended to any length.  Turning widgets for these paths also exist as we describe later.

\begin{figure}[h]
	\centering
	\includegraphics[scale = .25]{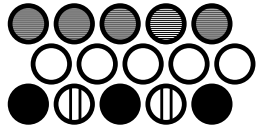}
	\caption{A Two-Symbol Path}
	\label{fig:twoPathWidget}
\end{figure}

The other type of path supports a chain of one of two different symbols.  In this type of path, shown in Figure \ref{fig:twoPathWidget}, whichever symbol is played forces the next play to follow suit.  If a space is barred, the next play must also be bars.  The same is true for filled (we assume again that the flow of play is going from left to right).

This type of path is often the by-product of play leaving other widgets.  Since all our widgets use single-symbol paths leading in, it is vital to be able to force the path to switch from a two-symbol path to a single-symbol path.  

\begin{figure}[h]
	\centering
	\includegraphics[scale = .25]{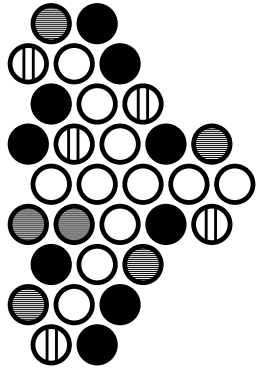}
	\caption{A Switch from Two-Symbols to Single-Symbol}
	\label{fig:twoToOnePathWidget}
\end{figure}

Figure \ref{fig:twoToOnePathWidget} shows a mechanism for this switch.  The prescribed flow of play in this widget in the diagram is again from left to right, strictly horizontal, although there are free spaces both above and below.  Indeed, if a player deviates by playing above the horizontal line, there is a winning response, simply by playing further above.  The same is true for playing below.

The beginning pattern of the single-symbol path is clear on the right-hand side of the widget, and, as before, no matter which play is made approaching that pattern, a sequence of fill plays can and must be made.

\begin{figure}[h]
	\centering
	\includegraphics[scale = .25]{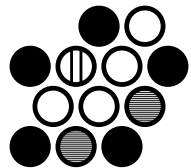}
	\caption{A 60 degree turn in a One-Symbol Path.}
	\label{fig:60TurnWidget1}
\end{figure}

\begin{figure}[h]
	\centering
	\includegraphics[scale = .25]{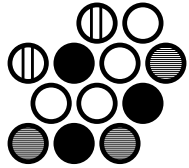}
	\caption{Another 60 degree turn in a One-Symbol Path.}
	\label{fig:60TurnWidget2}
\end{figure}

We must also be able to turn our paths in order to have them line up with other widgets.  Figures \ref{fig:60TurnWidget1} and \ref{fig:60TurnWidget2} reveal 60-degree turning options.  In order to turn further than 60-degrees, we can just pair two or three of these together to attain 120 or 180 degree rotations.  Note that in the second example (Figure \ref{fig:60TurnWidget2}) there are two possibilities for playing at the ``elbow'' of the turn.  This does not affect the overall restriction; further plays after the elbow must return to the original symbol.

Unfortunately for fans of two-symbol paths, we do not bother to create turning widgets for them.  Instead, whenever we are presented with a situation where a two-symbol path occurs, we will immediately switch it to a one-symbol path.  This does not present a problem, as only one of our widgets results in an out-going two-symbol path: the variable widget.  The out-going paths for these widgets will be followed by the two-to-one symbol switch widget.

\subsection{Variables}

\begin{figure}[h]
	\centering
	\includegraphics[scale = .25]{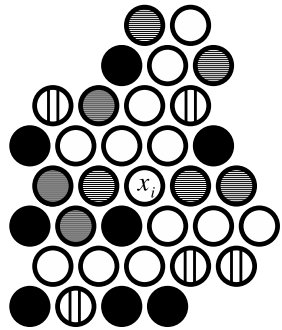}
	\caption{Variable Widget}
	\label{fig:variableWidget}
\end{figure}

Having described these devices, we are prepared to reveal the widget for modelling variables, presented in Figure \ref{fig:variableWidget}.

Here the flow of play enters initially from the bottom left and exits through the lower right path.  During this time, the ``playability'' of the circle corresponding to some boolean variable $x_i$ is determined.  If that variable is investigated at the end of the game, then the flow of play will enter through the entrance in the upper right corner, and will terminate inside the widget.  

The choice of symbol played in the space directly below $x_i$ determines the playability of $x_i$ (and corresponds to the assignment of true or false).  Since the plays up to that point must all either be fills, or---in the case of the last space---bars, the deciding play must also either be a fill or bars, and can be either, independent of what the previous play was.  Notice now that if a fill is made, the location $x_i$ will be playable later, whereas a play of bars means that $x_i$ is unplayable.  Thus, we associate a play of fill as setting $x_i$ to true, while a play of bars sets $x_i$ to false.
 
After this choice is made, the prescribed flow of play continues rightward, which is clearly forced in the case that $x_i$ is made unplayable.  We now must describe a winning response strategy to a player deviating from the play flow, which occurs when $x_i$ is played prematurely.

\begin{figure}[h]
	\centering
	\includegraphics[scale = .25]{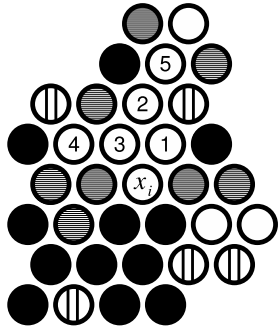}
	\caption{Responding to a Premature Play at $x_i$}
	\label{fig:prematureVariablePlay}
\end{figure}

\begin{lemma}
Prematurely playing at $x_i$ is suboptimal.
\end{lemma}

\begin{proof}
In order to prove this lemma, we must show a winning counter-strategy to a premature play in $x_i$.  We use the numbers from Figure \ref{fig:prematureVariablePlay} to refer to play locations in this strategy.  

Assume that an offending player played prematurely at $x_i$.  If they have not already lost, then they must have either shaded or filled $x_i$.  In either case, the winning punishment strategy begins by filling location $1$.  The offending player will have to respond by either

\begin{itemize}

\item[a)] Filling location $2$.  A winning response to this play consists of filling $3$.  Now, the offending player must play at the unplayable location $4$.

\item[b)] Filling or shading location $3$.  The winning response to this is to fill $2$.  Now, the offending player can only play at $5$, which is unplayable.

\end{itemize}

Thus, playing at $x_i$ prematurely is a losing strategy, and is suboptimal.

\end{proof}

Now, assuming the players follow the prescribed flow, the parity of playable spaces in the upper portion of the widget is defined by the playability of $x_i$.  If $x_i$ is investigated, then the flow of play will enter at the upper right from a single-symbol shading path.  Notice that these incoming plays cause location $1$ in Figure \ref{fig:prematureVariablePlay} to be unplayable.  Thus, the following sequence of non-suicidal plays is at locations $2$, $3$, then---if it is playable---$x_i$.  

Thus, if $x_i$ is playable, the play at $3$ loses (a play at $x_i$ forces the loss at $1$).  Otherwise, the play at $3$ wins, because all three neighboring spaces are unplayable.

\subsection{Splitting and Rejoining Paths}

When determining which variable to investigate, players need to be able to make choices to follow different paths.  In turn, multiple paths must converge towards the same variable, as multiple clauses can contain the same literal from our instance of \texttt{QSAT}.  Thus, we require widgets for splitting and rejoining paths.

In the course of any game, exactly one path through the variable investigation process will be used (unless a player performs obviously suboptimal moves beforehand).  We will enforce this through the design of these splitting and rejoining widgets, as well as the following crossover widget.

\begin{figure}[h]
	\centering
	\includegraphics[scale = .25]{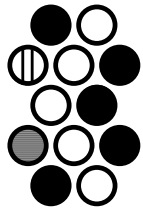}
	\caption{This widget splits a path.}
	\label{fig:splitterWidget}
\end{figure}

The splitter shown in Figure \ref{fig:splitterWidget} is extremely simple to comprehend: the flow of play goes from left to right.  Whichever player makes the second play in the widget has the ability to choose between the two paths.  Assuming the first player is forced to fill the first circle (by using a single-symbol fill path to enter this widget) the second player can choose to either take the upper or lower path.  The first player must fill the following circle, and play continues through a single-symbol path.

\begin{figure}[h]
	\centering
	\includegraphics[scale = .25]{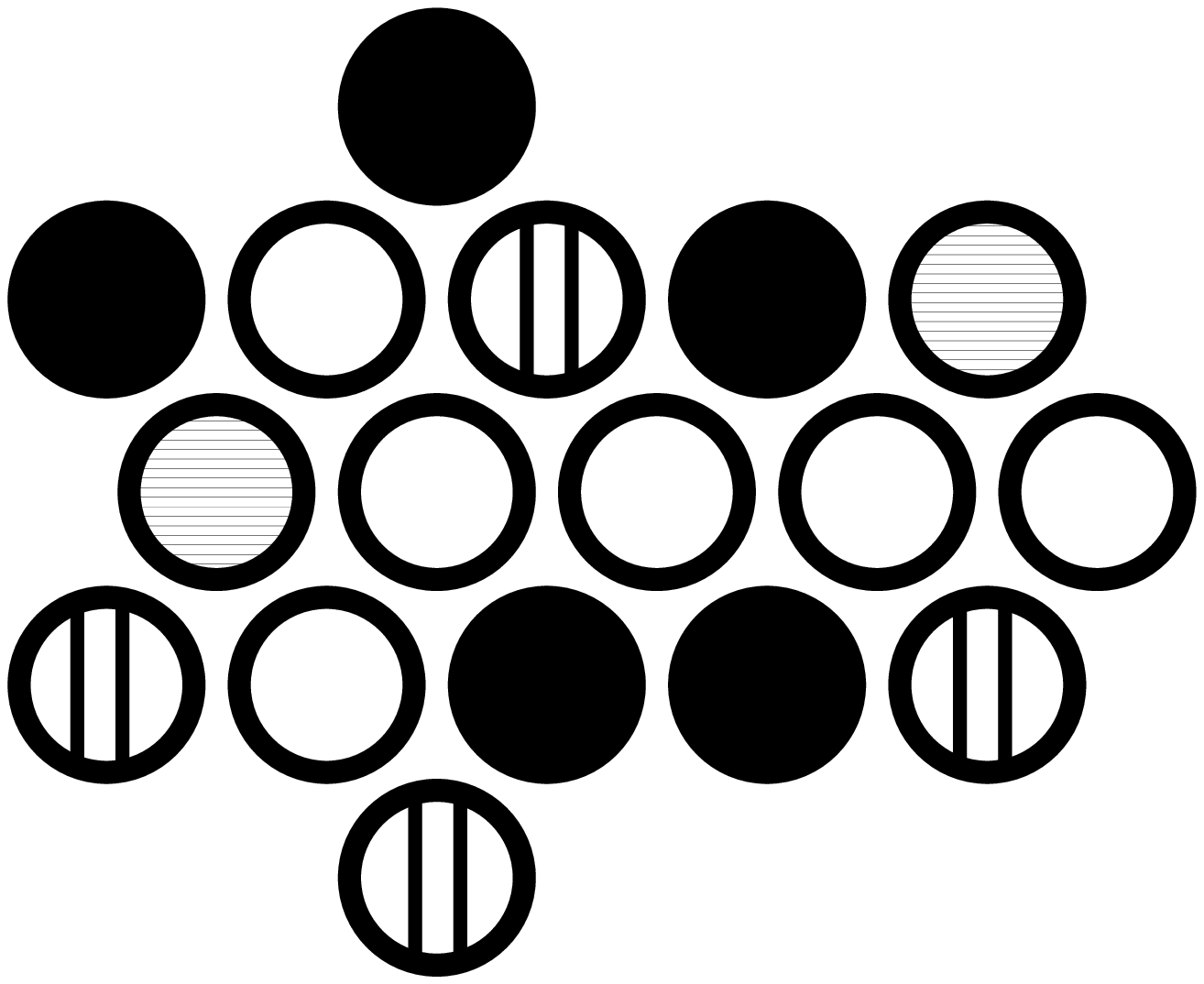}
	\caption{This widget converges two paths.}
	\label{fig:joinerWidget}
\end{figure}

Joiner widgets are slightly more complex, as we have to protect the flow of play (again, left to right in Figure \ref{fig:joinerWidget}) and prevent a player from ``going'' backwards.   The widget performs this automatically.  Notice that play from the top must begin (in the widget) with a fill, forcing the second play to be a fill.  Now, the space below is unplayable, preventing backtracking, and the players must continue to the right.  The same phenomenon occurs when play enters from below: forced bars cause the upper entrance to be unplayable.  Thus, our joiner widget ``works'' in the sense that it protects the flow of play against backtracking.

\subsection{Path Crossing}

\begin{figure}[h]
	\centering
	\includegraphics[scale = .25]{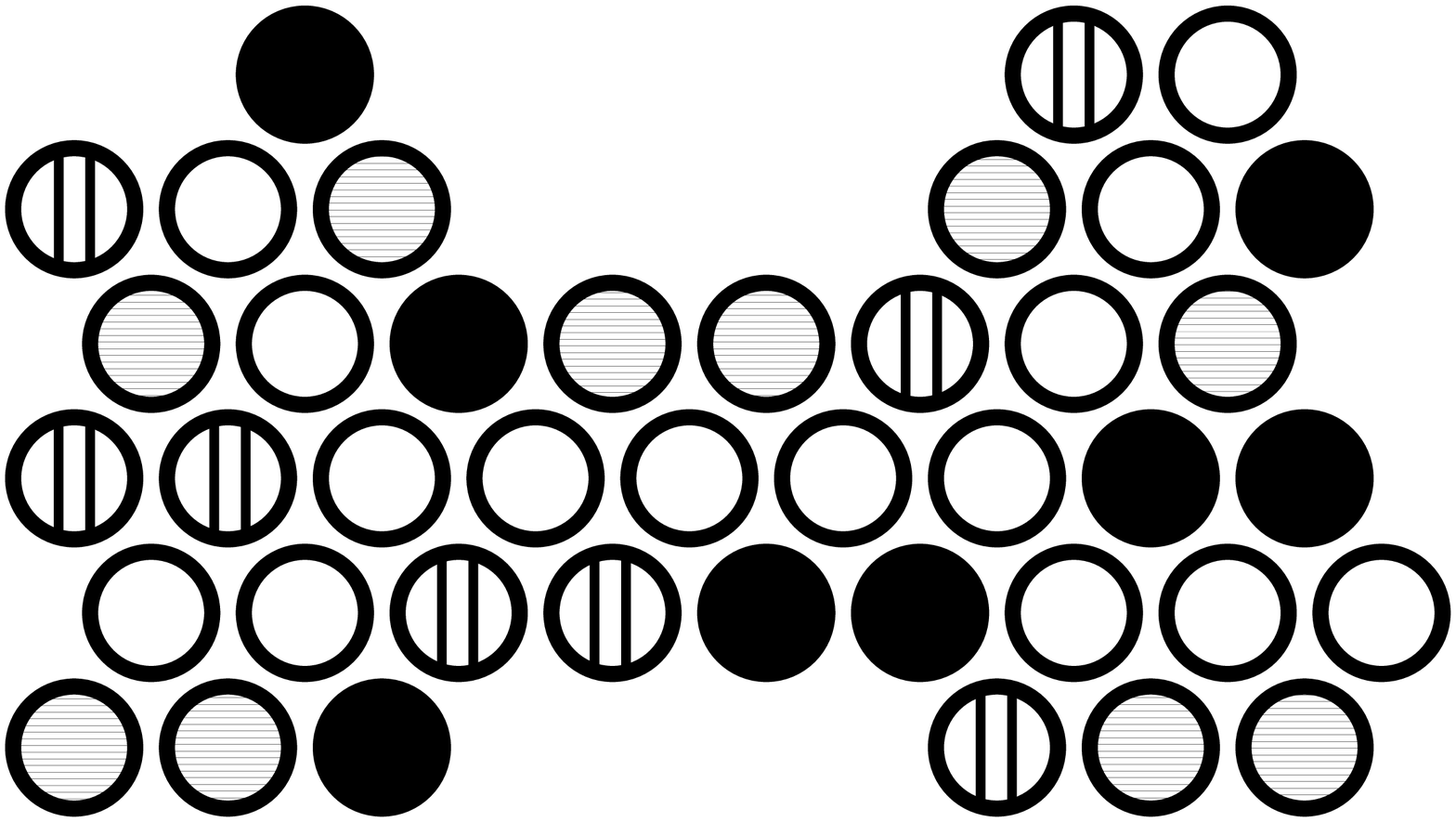}
	\caption{This widget crosses two paths.}
	\label{fig:crossoverWidget}
\end{figure}

The graph of paths in our model is not necessarily planar, meaning that we have to be able to handle paths which cross.  Thus, we need another widget that acts as a gate, forcing a crossing of two seperate possible play flows.  Luckily, such a widget exists, as pictured in Figure \ref{fig:crossoverWidget}.  Since play will follow exactly one path on the way to investigating a variable, each crossover piece will be used at most once during the course of a game.

We keep with the practice of using left-to-right play flow in Figure \ref{fig:crossoverWidget}, except that here we have two different possible flows: bottom-to-top and top-to-bottom.  We will follow one of the path flows and list the sequence of necessary and optional plays, showing that it ``flips'' sides while passing through the widget.  The reader may then investigate play flow from the other entrance.

Assume play enters the lower left side from a single-symbol fill path.  Then, the first two plays in the widget diagram must be fills, followed by either a fill or bars at the path intersections.  Either of these plays both makes a play upwards unplayable, and forces the next two plays to the right to be fills.  The next play must be a shading, followed by another shading in the intersection.  Now the lower path is unplayable, and play must continue upwards.  By verifying entrance from the upper left path, the reader will see that the crossover widget accomplishes its goal.

In truth, the widget is even more robust than we describe above.  As it turns out, threads of play may enter  from any adjacent two of the four sides, and the path-crossing property will still hold.  In our example construction in Figure \ref{fig:exampleWidgetsTogether}, play enters from either of the bottom holes and exits from one of the top holes (the example uses an upside-down version of the widget).  

\subsection{Parity Switching}

Finally, in order to ensure that the correct player has the chance to play at the variable locations $x_i$ during investigation, we may need to increase the lengths of some paths without shifting other widgets.  For each different possible path, we need to measure the length of the path to see which player would be playing on the variable site.

\begin{figure}[h]
	\centering
	\includegraphics[scale = .25]{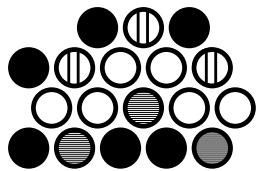}
	\caption{This widget effectively switches the order of play in a path.}
	\label{fig:paritySwitchWidget}
\end{figure}

If the literal in that clause is negated, then the adversary should be set to play on the variable.  Otherwise, in order to be consistent with the assignment of truth value to $x_i,$ it should be the hero's move that lands on the variable site.  Thus, for each path, we check the parity of the path length.  If it does not match the possible negation of the corresponding literal, as detailed above, we add a parity switch along the path between the last splitter and first joiner.  The parity switch widget is detailed in Figure \ref{fig:paritySwitchWidget}.  This device simply inverts the final order of play that would occur in a basic single-symbol path of the same lateral length.  

\section{Putting the Pieces Together}

We have described all our widgets, as well as given mechanisms for them to all fit together.  We are now prepared to prove the following theorem:

\begin{theorem}
Given a \texttt{TQBF} instance $f$, we can construct a Sperner game state such that the first player has a winning strategy if and only if $f$ is true.
\end{theorem}

\begin{proof}

This proof follows the intuition behind the construction portrayed in Section \ref{sec:reductionOutline}.

For each variable in $f$ we will use exactly one variable widget.  After each of these widgets, we will include a switch widget from a two-symbol path to a one-symbol path.  Following that widget, a path will connect to the entrance of the next variable widget.  We provide appropriate path lengths so that the same player will set all of the widgets corresponding to existential variables in $f$ while the other player will set all those corresponding to universal variables.  

After the last variable widget and its following path switch, we continue the path by splitting it into new paths corresponding to the clauses of $f$ with splitting widgets.  We create a tree structure such that at each split, the same player who sets the universal variables will choose the path.  

Once we have a path for each clause, we have a second round of splitting.  Here the player who sets the existential variables will choose between the three literals that exist in that clause.  Thus, after this splitting we will have three paths for each clause in $f$, or one for each appearance of a literal in $f$.

At this point, we will rejoin those paths corresponding to the same literal.  Thus, we will employ crossover widgets to route paths such that all paths corresponding to the same literal are adjacent.  Still before rejoining, we want to make sure that those paths whose literals are negated have a different path-length parity than those who are not.  Thus, we use parity switching widgets where appropriate so that this concern is met once the paths are joined.  Finally putting the rejoining widgets in place, we conglomerate all the paths corresponding to the same literal.

In the final step, we reconnect these paths to the corresponding variable widgets, through the entrance shown at the top of Figure \ref{fig:variableWidget}.  We may need to add a final parity switch here to assure that paths corresponding to unnegated literals predict that the existential player will color the circle corresponding to the variable setting and that the universal player will play on those corresponding to negated literals.  

Finally, before the first variable widget, we need to include a starting circle.  In order to do this, we choose a circle near the first variable and build a path to that widget such that the player coloring the starting circle will also set the first variable.  Then, before the starting circle, we color in the circle to its left, declaring it the ``last-played'' circle.  We color all neighbors of that last-played location, aside from the starting circle, leaving the path intact.  Once this is in place, the widgets have been correctly assembled.  It remains to be shown only that this is a legitimate game state.

In order to do this, we must assert that our construction both fits within an initial Sperner triangle, and also can be constructed from a legal series of moves.

First we investigate fitting our construction into a Sperner triangle.  This is an easy obstacle to overcome: we can simply draw a large triangle around our game board.  We can color as many circles as necessary outside our structure, so long as we don't create a bad triangle.  

To address the second concern, it is true that not every semi-coloring of circles is attainable.  For instance, the board in Figure \ref{fig:unrealizableBoard} cannot be realized through a playing of the game.  We quickly make the following definition:

\begin{definition}
An \emph{island} is a connected collection of colored circles surrounded by circles which are each either uncolored or a part of the boundary.
\end{definition}

The board in Figure \ref{fig:unrealizableBoard} contains two unconnected colored ``islands'', neither of which has a place in which a player might have been allowed to jump.  Thus, islands can exist only if they contain a colored circle completely surrounded by other colored circles, or if the last play was made within the island.  Otherwise, since players must play adjacent to the last play whenever possible, there is no way for play to have jumped from one island to another.  Therefore, all but one island must contain a colored circle surrounded by colored circles to represent the location from which a player was allowed to jump.  

\begin{figure}[h]
	\centering
	\includegraphics[scale = .25]{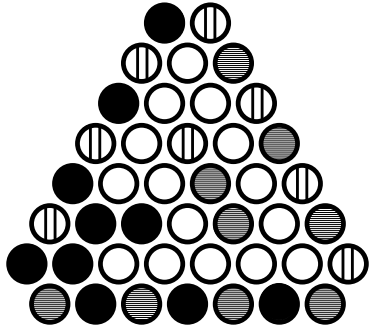}
	\caption{This board contains two disconnected islands which cannot co-exist.}
	\label{fig:unrealizableBoard}
\end{figure}

More formally, we can define the legality of an island:

\begin{definition}
An island is \emph{legal} if it can be constructed from a sequence of disjoint paths, each of which ends in a colored circle which is either the last play made, or is surrounded by colored circles from this or previous paths.
\end{definition}

Ensuring that jumps outside could have occured is simply a problem of providing enough space between our widgets.  Indeed, we only need a hexagonal structure of seven colored circles in order to construct a legal island; our sequence of plays can first color the six circles on the outside of the hexagon, then color the middle circle.  Thus, since we only need a little space between widgets to ensure possible jumps, it is a simple matter to ensure that the holes between paths in the reduction are large enough.  If at any point one of the holes is too small, we simply increase the distance between widgets by extending the paths surrounding it.

\end{proof}

We continue by providing a sample reduction to \texttt{SPERNER} from a \texttt{TQBF} instance.

\subsection{Sample Reduction}

We steal our example \texttt{TQBF} instance from a \texttt{GEOGRAPHY} example of Papadimitriou's \cite{PapadimitriouBook:1994}: $\exists x: \forall y: \exists z: \left[ (\overline{x} \vee \overline{y}) \wedge (y \vee z) \wedge (y \vee \overline{z})\right]$.  Although this example contains only two literals per clause, the construction utilizes all of our widgets and correctly portrays the ease with which they fit together.

\begin{figure}[h]
	\centering
	\includegraphics[scale = .60]{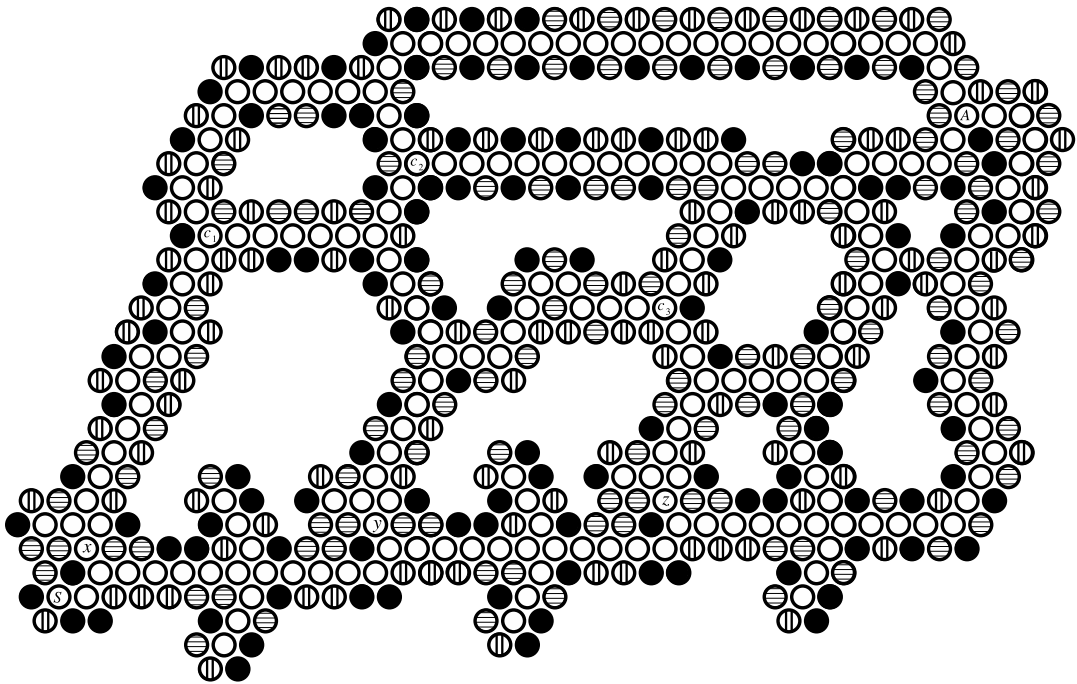}
	\caption{Widget assembly for $\exists x: \forall y: \exists z: \left[ (\overline{x} \vee \overline{y}) \wedge (y \vee z) \wedge (y \vee \overline{z})\right]$}
	\label{fig:exampleWidgetsTogether}
\end{figure}

In Figure \ref{fig:exampleWidgetsTogether}, we have assembled the board using our widgets.  We use $S$ to denote the starting position (we assume that the adversary last colored the circle directly to the left of the $S$ and that no other uncolored nodes are adjacent to that circle).  Some of the other nodes have also been marked to show their correspondence to the formula.  Each of the variables $x,$ $y$ and $z$ are noted in the Figure, as well as the clauses $c_1 = \overline{x} \vee \overline{y},$ $c_2 = y \vee z$ and $c_3 = y \vee \overline{z}$.  The $c_i$'s indicate locations where the hero will choose between one of the literals in that clause.  In addition, we have marked the split $A$ where the adversary makes their first choice between clauses.  If they wish to investigate $c_1$ or $c_2$, then they must take the upper path.  Otherwise, choosing the lower path will lead directly to $c_3$.  

The attentive reader can count spaces to see that the hero will choose to set $x$ and $z$ and will get to choose between literals at each clause.  The adversary gets to set $y$ as well as choose between clauses.  Furthermore, after each literal is chosen from each clause, the reader can count to see that the adversary is scheduled to play on the variables when they are negated in the appropriate clause, while the hero is scheduled to play on them when they remain un-negated. 

\begin{figure}[h]
	\centering
	\includegraphics[scale = .60]{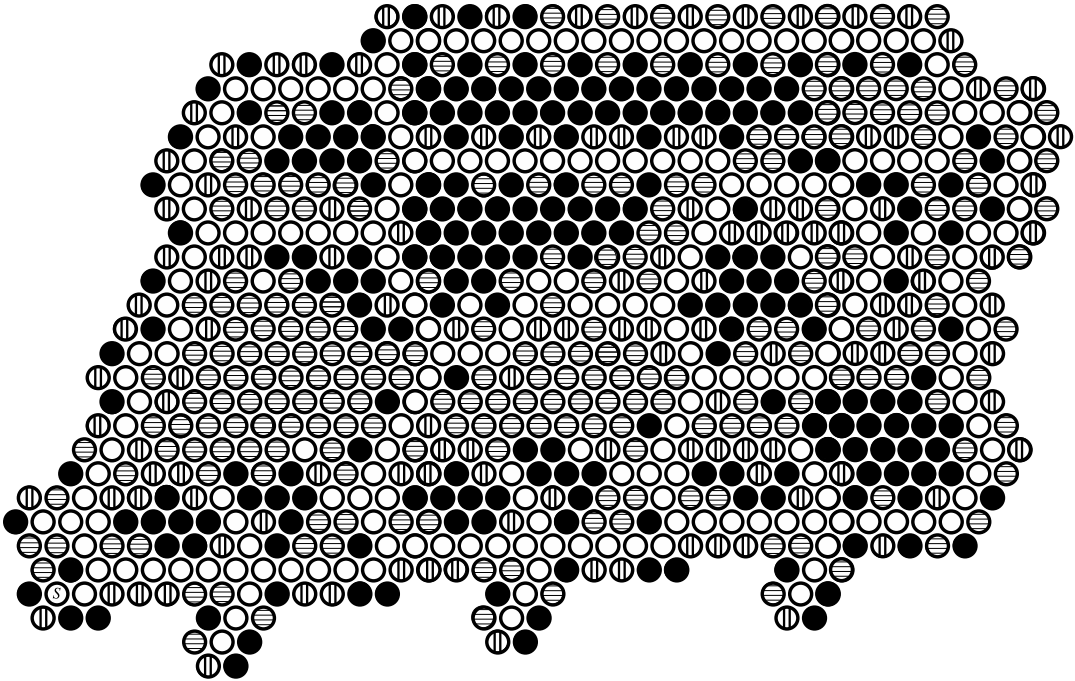}
	\caption{Filling in Holes Between the Widgets.  We also remove the unnecessary notation from the nodes ($S$ is still the start node, and must be denoted).}
	\label{fig:exampleReductionBoard}
\end{figure}

Demonstrating that this is indeed a legal game state, Figure \ref{fig:exampleReductionBoard} shows the holes of our diagram filled in.  Recalling earlier discussion (and Figure \ref{fig:unrealizableBoard}), these holes must each simply contain a hexagon of seven colored circles connected to the border.  We go a bit further here, and color all circles in the holes, aside from some doomed ones, although this is not necessary to the process.

Having removed all unnecessary notation from the game board, we must just imagine that this structure exists within our Sperner Triangle.  Thus, we must only implant our construction inside a large-enough initial board.

\section{An Open Question and a Conjecture}
\label{sec:unrestrictedSperner}
The unrestricted Sperner game could be an interesting
  variation of Sperner for beginners.
It is in fact an amusing game and can be useful
 to gain intuition about the parity
 underlying the Sperner's Lemma without attending
 Topology classes.
Although its complexity is still open, we conjecture that
 this parity structure will lead to a polynomial-time
 solution.

We would like conclude our discussion about Sperner by outlining a
 possible approach to solve this open question.
In this approach, the winning strategy for playing
 unrestricted Sperner revolves around a simple
 relation of bad triangles to unplayable nodes.
If, at the end of the game, any losing move will
 create exactly one (or an odd number of)
 bad triangle(s), then we already know who should
 lose the game, depending on the number of
 uncolored circles on the initial triangular gameboard.

Assuming there will be an odd
 number of doomed circles at the end of the game,
 just before the last player is forced to make a losing play,
 the parity of the number of initial circles determines the winner of
 the game.
If the board begins with an odd number of spaces, the first
 player will lose.
Otherwise, the second player will be on track to
 lose the game.
The only strategy remaining for this second player is
 to somehow force one of the losing plays
 to create two (or an even number greater than zero) bad triangles.
This causes a sort of
 ``parity switch'', resulting in an even number of doomed circles.

We conjecture that due to the fragile structure
 of constructs that result in these parity
 switches, this circumstance is easily avoidable.
The player initially winning as predicted above can
 prevent parity switches from occuring in most situations.
In fact, we would like to show that this player needs to
 look ahead only a constant number of turns in the game tree
 to avoid traps leading to  parity switches.

If this conjecture is true, then it is easy to determine
 who has a winning strategy at any point in time,
 simply by investigating the parity.
The key to prove this conjecture is to
 show that the supposed winning player needs
 only to keep track of the possible
 formation of a small number parity-switching cases
 and show that the recognization of these cases can be
 performed in time polynomial in the
 number of open spaces on the gameboard.
In doing so, we could prove that strategies for this game are in $P$.

\section{More Fixed-Point Games}

The principle of our game on Sperner triangles can be applied to
design other Fixed-Point Games. Each of them is based on a discrete
fixed-point theorem.
Perhaps, closest to our game on Sperner triangles is the game defined
on grid $[1:n]^2$ based on the following fixed-point theorem.

\begin{theorem}[\cite{DAS:05,CDT:06,CHE:06}]
Suppose vertices of a 2D grid over $[1:n]^2$ are colored according to the
following rule:
Vertices on the left boundary are filled, vertices on the bottom
boundary, other than the one that is also on the left-boundary, are
shaded, the remaining boundary vertices are barred and the interior vertices
are arbitrarily colored. Then, there must be a unit square that
contains all three colors.
\end{theorem}

The game board is then the two-dimensional $n$ by $n$ grid. The
boundary of the board is colored according to the theorem above. At
each odd step, the hero colors a vertex and at each even step, the adversary
colors a vertex. The player that generates a tri-chromatic unit square
loses. As a decision problem, given a legal configuration of the game
board, we need to determine whether the hero has a winning strategy.

Similarly, we can define two versions of the game: The restrictive
game requires that if possible, the chosen vertex is adjacent to the
vertex last colored. We call this game \emph{2-D-Brouwer}.  In the unrestrictive game, players can color any
uncolored vertex. We anticipate that our $PSPACE$-complete result on
Sperner game can be extended to 2-D-Brouwer.

Another interesting discrete fixed-point theorem is the following
discrete Brouwer fixed-point theorem. A function $f: [1:n]^2
\rightarrow \{(0,0), (1,0), (0,1)\}$ is bounded if $f(x) + x \in
[1:n]^2$ for all $x \in [1:n]^2$ and it is direction preserving if
$||f(x)-f(y)||_1 \leq 1$ for all pairs $x,y \in [0,1]^2$. A vertex
$v\in[1:n]^2$ is a zero-point of $f$ if
$f(v) = (0,0)$.

\begin{theorem}[\cite{IMT:2005}]
Every bounded and direction-preserving function over $[1:n]^2$ has a zero point.
\end{theorem}

The game board is again the two-dimensional $n$ by $n$ grid. At each
odd step, the hero chooses an unassigned vertex $v$ and assign to it a
vector from ${(0,0),(0,1),(1,0)}$ that does not violate the bounded and
direction preserving conditions. At each even step, the adversary does
the same. A player's move lead to a zero-point assignment loses the
game. We can similarly define the restrictive and unrestrictive
version of this game and the decision problem.  We call the restrictive version of this game \emph{Flow}.
Again, we anticipate that Flow is a $PSPACE$-complete game.

Our games from Sperner's Lemma might also inspire the creation and
study other games defined based on combinatorial problems that might
have a pure Nash equilibrium. For example, consider the following. We
are given a directed graph where each vertex has $k$-directed links.
At each odd step, the hero chooses a vertex $v$ and rewires its out-links
to strictly improve a quality measure of the resulting graph.  This measure could be, for instance, the total
distance or maximum distance of $v$ to other vertices. At each even step, the
adversary does the same. Each of these moves leads to a pure Nash
equilibrium, namely, where no vertex has a rewiring which strictly improves the measure.  Once this equilibrium is reached, the game ends with the last-improving player as either the winner or loser (depending on the rules).
The decision problem could be, given a strongly connected graph in
which each vertex has degree $k$, to determine whether the hero has a winning
strategy. 

\section{Conclusion}

With the recent amount of attention paid to the implications of Sperner's Lemma in the theoretical computer science community, the study of Sperner and other similar fixed-point games might enhance our understanding of the complexity of fixed-point computation.  Indeed, with the newfound relationship between the complexity class $PPAD$, fixed points and Nash equilibria \cite{Papadimitriou:1994,DAS:05,CDT:06,CHE:06}, this is a promising avenue for continuing study.  

Assuming our conjecture for the unrestricted version of Sperner holds, we would have already generated an environment which somehow contains the boundary between $P$ and $PSPACE$ (should one exist).  The solution of our open question could lead immediately to another: How much distance do we allow between plays before the game is no longer $PSPACE$-complete?  For example, what happens to the complexity if we allow one player to color in a circle with distance \emph{two} from the last-colored circle?  What if we allow a non-constant amount of space between plays?  Answers to these questions may reveal additional information about the divisive properties of complexity classes.

Aside from the ramifications of our conjecture, Sperner is a simple, $PSPACE$-complete game rising from a mathematically-significant construct.  It inherently carries a fixed-point awareness (a loss in the game corresponds to the creation of a fixed-point) and simultaneously avoids the first-player-wins dilemma faced by other related games.  Most exciting, perhaps, is that it accomplishes both these benchmarks without sacrificing any elegance.  

\begin{figure}[h]
	\centering
	\includegraphics[scale = .60]{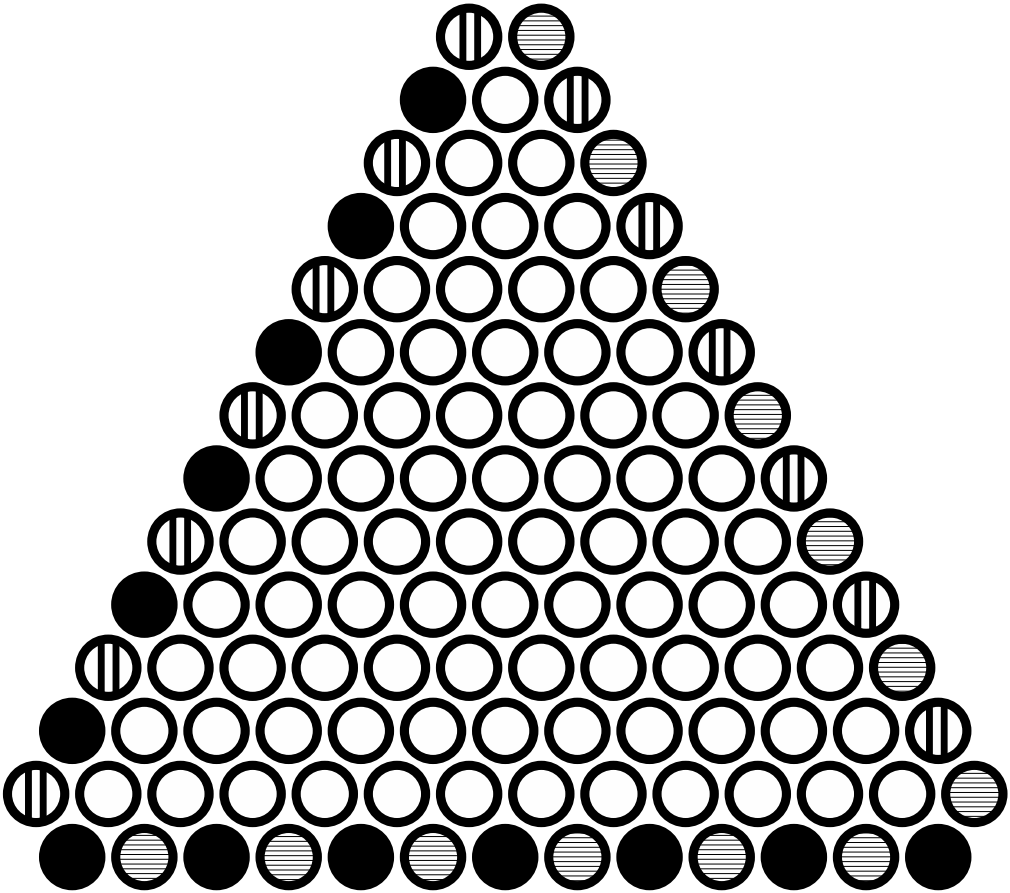}
	\caption{A board of size 12.  Boards of other sizes are accessible for play at {\it http://cs-people.bu.edu/paithan/spernerGame/}}
	\label{fig:sampleboard12}
\end{figure}

For beginning players, we suggest starting on a game board with seven circles on a side.  In the spirit of marketing the game at a specific size (Hex is defined on an 11 x 11 hexagonal grid, while Go is historically played on a 19 x 19 grid) we suggest that serious players challenge each other on a board with 12 circles on a side.  We provide a sample board of size 12 in Figure \ref{fig:sampleboard12} (on the last page) to facilitate this competition.

\section{Acknowledgments}

Kyle wishes to thanks all his friends at Boston University, his mother Janelle, Silvia Montarani and Katelyn Mann for playing Sperner with him.
We also thank Xi Chen for teaching us many cool fixed-point theorems and Scott Russell for proofreading our paper.

\bibliographystyle{plain}
\bibliography{paithan}

\clearpage

\vspace*{1.5in}

\end{document}